\documentclass[authoryear,preprint,12pt,3p]{elsarticle}
\usepackage[utf8]{inputenc}
\usepackage{t1enc}
\usepackage[dvipsnames]{xcolor}
\usepackage{float}
\usepackage{comment}
\usepackage{subcaption}

\usepackage{amssymb}

\biboptions{comma,round,sort}

\journal{Icarus}

\begin{document}

\begin{frontmatter}

\title{Annual and daily ideal periods for deliquescence at the landing site of InSight based on GCM model calculations}

\author[label1,label2]{Bernadett P\'al\corref{cor1}}
\address[label1]{Konkoly Thege Miklos Astronomical Institute, Research Centre for Astronomy and Earth Sciences, MTA Centre of Excellence, Budapest, Hungary}
\address[label2]{Eötvös Loránd University, Budapest, Hungary}

\ead{pal.bernadett@csfk.mta.hu}

\author[label1]{\'Akos Kereszturi}

\begin{abstract}
Liquid water is one of the key elements in the search for possible life outside of the Earth and has a wide range of consequences on various chemical and geological processes. The InSight probe landed on Mars with a special equipment dedicated to examine geophysical characteristics and internal heat flow of the planet and some meteorological instruments also included in the payload. We examine the annual and daily variations of near-surface relative humidity and surface temperature calculated from the General Circulation Model (GCM) at Elysium Planitia, the landing site of InSight and search for possible ideal times for deliquescence. We inspect three different hygroscopic salts, but find that out of the three only calcium-perchlorate could liquify at the environment of InSight. We find that nighttime ideal periods could occur in a limited window between approximately Ls 90$^{\circ}$ and 150$^{\circ}$ at the late evening hours centered around 9 PM. In our daily studies we find no instances where the whole night could be ideal for deliquescence. This is mostly due to the temperatures dropping below eutectic level leading to a 0.5 - 2 hour long presumed ideal period before midnight. On multiple occasions the temperature is just a few degrees below the necessary limit while relative humidity is high enough, therefore the precise temperature measurements of InSight could be critical in determining ideal periods for deliquescence. \\
\end{abstract}


\end{frontmatter}


\section{Introduction}
\label{sec:intro}
Our aim is to investigate the potential of deliquescence at the landing site of InSight mission by the related conditions and processes to see how it could influence the hypothesis of nighttime microscopic liquid layer formation. The focus of this work is to find seasonal and daily periods, when the conditions could be favorable for deliquescence and provide context to compare the possibility of this process to other Mars surface locations. This work could also point to seasonal and daily periods, which should be analyzed in detail in the high resolution meteorological dataset of InSight to support the further improvement of climate models --- to forecast periods favourable for the appearance of liquid phases. \\  

The landing site of the InSight mission is at Elysium Planitia, located at 4.5$^\circ$N 135.9$^\circ$E, which was chosen after considering both the engineering and scientific constraints. Low latitude spot for strong solar illumination, low elevation for effective parachute driven deceleration and a smooth, flat surface with low rock coverage for safe landing and uncomplicated access to the subsurface \citep{golombek2017,golombek2016}. The area is around the Early Hesperian transition unit \citep{tanaka2014}, covered by the Elysium Mons volcanits \citep{vaucher2009}. Based on the analysis of rocky ejecta craters, the landing site is covered by at least 3 m thick regolith \citep{warner2017}. Phyllosilicate was also identified nearby at the southwest flank of Elysium Mons inside a crater \citep{pan2018}, which might be the exposed part of a subsurface layered unit, where a clay-bearing sedimentary unit underlies the thick lava flows. The landing site is mainly covered by atmospheric dust representing the globally wind homogenized material on Mars. Its composition is expected to be weathered basaltic with high Fe/Mg, low Al$_2$O$_3$, made up of plagioclase, pyroxenes and olivine minerals \citep{riu2018}. It is enriched in Ti, Cr, Fe, S, and Cl \citep{berger2016} with Fe-rich amorphous composition. The material making up the regolith is dominated by palagonization process \citep{mcsween2000} and evaporation \citep{elsenousy2015}, as well as atmospheric generated oxidant deposition \citep{atreya2006}. Although chlorine seems to be widespread in the regolith of Mars, the exact form of chemical occurrence is unknown. However, as it was identified in the form of perchlorates at high latitudes (by Phoenix) and low latitudes (by Curiosity) as well, it seems to exist globally \citep{clark2016} and it could play a role in the process of deliquescence. \\

The wind might have significant effect on seismometry realized on Mars \citep{nakamura1979}, thus InSight carries wind (1 Hz sampling by two separately mounted booms), air temperature (1 Hz sampling, 5 K accuracy, 0.1 K resolution) and pressure (20Hz sampling, noise level about 5 mPa) sensors \citep{banfield2014}, providing the best continuous dataset as the most complete record at a landing site ever. The system called Temperature and Winds for InSight (TWINS) is based on the heritage of the Rover Environmental Monitoring Station (REMS) onboard Curiosity rover \citep{velasco2015} that includes a magnetometer too for possible observations related to the ionosphere \citep{yu2018,spiga2018}. \\

The debate of potential existence of liquid water on the surface of Mars is still ongoing. There are theoretical models predicting the presence of liquid water \citep{clow1987, hecht2002, haberle2001}, and some surface features favour the ephemeral appearance of a liquid phase \citep{kereszturi2009, knauth2002, brass1980, kossacki2004, Mellon2001, mcewen2014, motazedian2003, szynkiewicz2009}. The emergence of liquid water on a microscopic scale, especially brines are favoured by the computational researches \citep{kossacki2008, mohlmann2004, martinez2013}, with the presence of brines suggested by observations of the Phoenix landing site and other locations \citep{chevrier2009, renno2009, kossacki2004, hecht2009}. These brines may act as possible agents of the formation of some recent flow-like features on Mars. Recurring Slope Lineae \citep{ojha2015, mcewen2011} may also be connected to some form of liquid, however \citet{dundas2018} showed that they are largely consistent with the dust-avalanche model as well.  \\

The meteorological observations of the Curiosity rover indicated that the nighttime conditions are favourable for the emergence of a thin, microscopic liquid film on the surface of hygroscopic mineral grains on the Martian surface and shallow subsurface \citep{torres2015}. Hygroscopic salts could adsorb water vapor directly from the atmosphere where deliquescence (transition from solid to aqueous phase), or efflorescence (transition from aqueous to solid phase) happen \citep{gough2011}. The threshold values of these processes are the eutectic relative humidity (RH$_{eut}$, where the deliquescence process starts), the deliquescence relative humidity (DRH, where the deliquescence process is complete) and efflorescence relative humidity (where the thin liquid layer is lost and the material turns into a solid phase). Deliquescence phase transition occurs when the local relative humidity (RH) is equal to or exceeds the DRH level. \citet{rivera2018} showed that there is a third and fairy restrictive criteria, that the saturation with respect to ice should not greatly exceed 1, thus we also take that into consideration when selecting the proposed ideal times.  \\

\section{Methods}
\label{sec:methods}

In determining the ideal periods we used the eutectic temperature and water activity of a solution at the eutectic temperature values detailed in Table \ref{tab:salts}. The values of calcium-perchlorate (Ca(ClO$_4$)$_2$) are from the work of \citet{toner2014}, the magnesium-perchlorate (Mg(ClO$_4$)$_2$) from \citet{mohlmann2011} and the calcium-chloride (CaCl$_2$) from \citet{davila2010}. Mg(ClO$_4$)$_2$ was identified by the Phoenix lander and is one of the most analyzed salts \citep{hecht2009} to date. The small spheroidal shaped features were observed by the robotic arm camera on the leg of the lander appeared to merge over time, which some scientist argue could be a sign that these features were in a liquid state \citep{renno2009}. It is possible, that these features were produced by liquid brine form of the aforementioned hygroscopic salts, however unfortunately detailed observations are not available from the acquired images. \\

\begin{table}[H]
    \centering
    \begin{tabular}{c c c} \\ \hline
       Salt  & eutectic temperature & Water activity  \\ \hline
       Ca(ClO$_4$)$_2$ & 199 K & 0.51 \\
       Mg(ClO$_4$)$_2$ & 212 K & 0.53 \\
       CaCl$_2$ & 223 K & 0.62 \\ \hline
    \end{tabular}
    \caption{Parameters of the examined hygroscopic salts and water activity of the eutectic solution.}
    \label{tab:salts}
\end{table}

We calculated the relative humidities for the Martian near-surface atmosphere from the data of the Laboratoire de M\'et\'eorologie Dynamique Mars General Circulation Model (LMDZ GCM), detailed in \citep{forget1999}, including a water cycle as described in \citep{navarro2014} for the whole Martian Year 29. The more recent validations of the model are detailed in \citet{millour2014} and \citet{millour2015}. This numbering of martian years is widely used and follows the calendar proposed by R. Todd Clancy \citep{clancy2000}, which begins on April 11, 1955 (Ls 0$^\circ$). The year 29 is without a global planet-encircling dust storm. To derive the relative humidity first the saturation water vapor volume mixing ratio is calculated with an equation used in the Martian Climate Database based on the Goff-Gratch equation \citep{goff1946, list1984} with respect to ice ($\mathrm{Q_{{sat}_i}}$) and with respect to liquid ($\mathrm{Q_{{sat}_l}}$):

\begin{equation}
    \label{eq:qsati}
    \mathrm{Q_{{sat}_i}} = \frac{100}{\mathrm{P}} \times 10^{2.07023 - 0.00320991 \mathrm{T} - \frac{2484.896}{\mathrm{T}} + 3.56654 \mathrm{log(T)}}
\end{equation}

\begin{equation}
    \label{eq:qsatl}
    \mathrm{Q_{{sat}_l}} = \frac{100}{\mathrm{P}} \times 10^{23.8319 - \frac{2948.964}{\mathrm{T}} - 5.028 \, \mathrm{log(T)} - 29810.16 \, \mathrm{exp}(-0.0699382 \mathrm{T}) + 25.21935 \, \mathrm{exp}(\frac{-2999.924}{\mathrm{T}})}
\end{equation}

\noindent where $\mathrm{P}$ is the surface pressure and $\mathrm{T}$ is the surface temperature. \\

\noindent After determining the necessary $\mathrm{Q_0}$ value, the relative humidity with respect to ice ($\mathrm{RH_i}$) or with respect to liquid ($\mathrm{RH_l}$) can be calculated as:

\begin{equation}
    \label{eq:rh}
    \mathrm{RH_{i,l}} = \frac{\mathrm{Q_0}}{\mathrm{Q_{sat_{i,l}}}}
\end{equation}

\noindent The water vapor volume mixing ratio (VMR) calculations are not reliable under approximately 4 meters above surface due to the complexity of near-surface interactions. By assuming the vapor to be well mixed between the surface and 4 meters, we can estimate the near-surface relative humidity levels. Because of the comparison of atmospheric vapor values with surface temperature and surface pressure, the relative humidity calculations with respect to ice can reach unrealistically high outlier values in a few instances. Physically the supersaturation never goes very high, because the nighttime atmosphere is much warmer than the surface, and just near it the atmosphere is drier due to surface interactions. In some way by assuming the water vapor values to be well mixed between the surface and approximately 4 meters, we show a ``potential supersaturation". To keep these outlier values from obstructing the visibility of the annual curves, we maximized the degree of supersaturation in 2. \\

We used Martian solar longitudes to mark the time of the year as illustrated in \citet{pal2019}, and indicated the times of day by using local Martian times in our figures. Please note, that the graphs showing daily variations go from 15:00 local time to next day 15:00 and not from 00:00 to 24:00. We chose to create the figures this way to place the night, the most interesting time of the day for deliquescence in the middle of the graph instead of cutting it in half. \\

We first calculated annual surface temperature and relative humidity values from the GCM model output at a given time of the Martian day. Comparing the values with the properties of hygroscopic salts (Table \ref{tab:salts}) we found that calcium-perchlorate would be the only one with a chance of deliquescence. After investigating the annual graphs we selected a number of solar longitude ranges when the circumstances could be ideal for liquefaction. In our next step we calculated ideal time periods at these Ls values throughout a Martian day and looked for similarities between the different solar longitude times. A selection of the graphs are shown in the results section with two figures at the end summarizing the outcome of the daily ideal time investigations. The figures included in this paper were picked out as to present general common features of all graphs investigated and to also show the not ideal instances. \\ 

\section{Results}
\label{sec:results}

In this section a selection of investigated graphs are shown with the annual curves in Subsection (\ref{subsec:annual}), the daily graphs selected according to the annual results in order to find the times of the highest chance of deliquescence (\ref{subsec:daily}) and the summarizing figures of ideal daily periods in subsection \ref{subsec:ideal}. \\

\subsection{Annual results}
\label{subsec:annual}

The resulting figures in this subsection show the annual variation of surface temperature and relative humidity at a given Martian local time. Deliquescence has a theoretical chance to occur if both the temperature and the relative humidity with respect to liquid are above a certain salt specific threshold at least at two subsequent data points. Furthermore we calculated the relative humidity with respect to ice as well to check if it does not exceed 1 by too much. If it exceeds by far, then according to \citet{rivera2018} water vapor should be nucleating ice instead of forming brines. Finding these ideal periods is a non-trivial task due to the overall Martian dryness and cold. Usually if the temperature is high enough the humidity is low and vice versa. This is rather apparent at all annual graphs. Usually at ideal times if the relative humidity is above the DRH of calcium-perchlorate, the surface temperature stays just above the eutectic temperature (ET) resulting in ideal spikes rather than lengthy consecutive periods in most cases. The presumed ideal periods are indicated by light green shaded rectangles in all figures. \\

\begin{figure}[H]
    \centering
    \includegraphics[width=\linewidth]{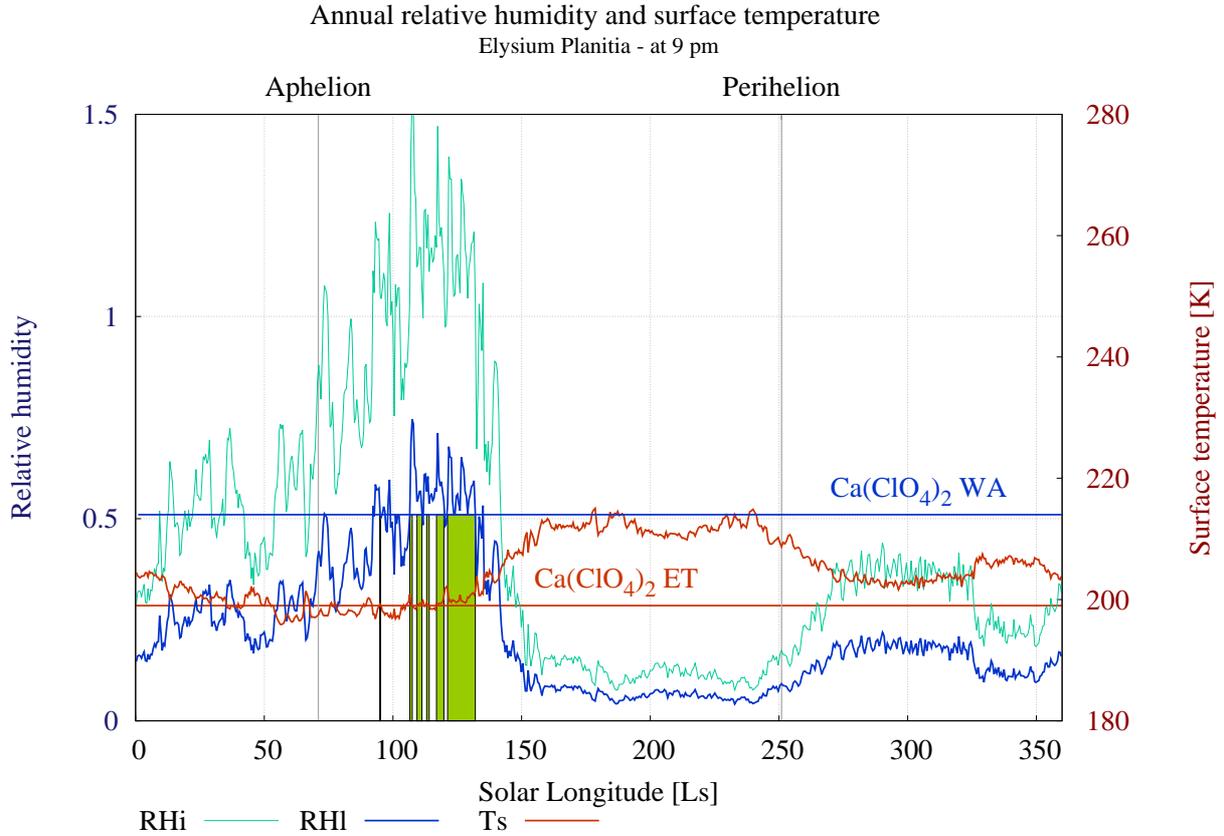}
    \caption{Annual variations of relative humidity and surface temperature at 9 PM local time at Elyisum Planitia. The light green shaded vertical rectangles show that the possible ideal times for Ca(ClO$_4$)$_2$ deliquescence occur only in a narrow time window between approximately Ls 90$ ^{\circ}$ - 140$ ^{\circ}$.}
    \label{fig:annual9pm}
\end{figure}

Figure \ref{fig:annual9pm} depicts the annual variation of surface temperature and relative humidity at 9 PM Martian local time. Looking at the horizontal blue and red lines indicating the minimum DRH and ET levels of calcium-perchlorate and the light green rectangles, one can see, that theoretically ideal times occur in the first third of the year only between approximately Ls 90$ ^{\circ}$ - 140$ ^{\circ}$. Apart from this time window, the relative humidity stays below the minimum necessary value of calcium-perchlorate, never reaching saturation. This results in the limited, but concentrated ideal regime regardless of the temperatures remaining over the eutectic level almost throughout the entire year. \\ 

\begin{figure}[H]
    \centering
    \includegraphics[width=\linewidth]{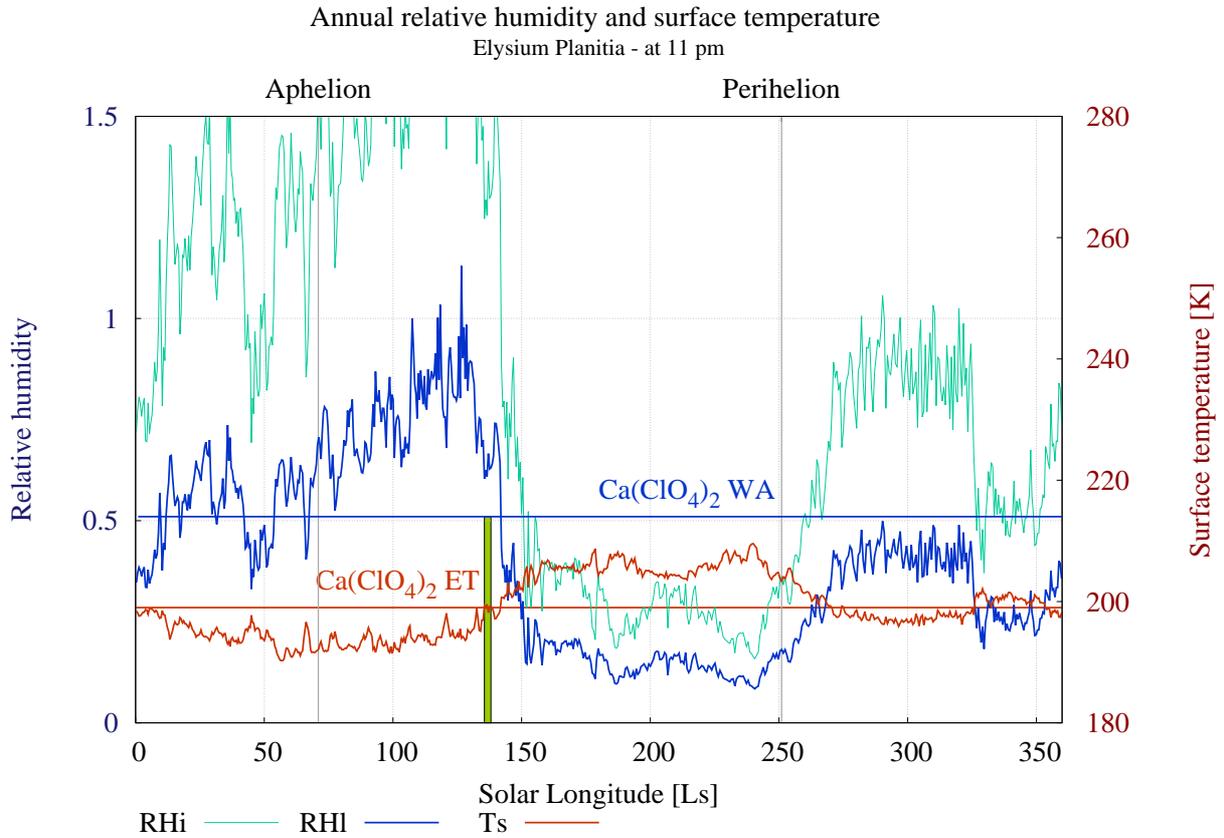}
    \caption{Annual variation of relative humidity and surface temperature at 11 PM local time at Elysium Planitia. The light green rectangles show that the predicted ideal times for Ca(ClO$_4$)$_2$ deliquescence might occur only briefly around Ls 140$ ^{\circ}$.}
    \label{fig:annual11pm}
\end{figure}

In Figure \ref{fig:annual11pm} one can see a different scenario compared to Figure \ref{fig:annual9pm}. While at 9 PM there was a short but almost continuous window around between the northern summer and autumn, at 11 PM the possibility of deliquescence seems rather slim. The relative humidity levels increase compared to 9 PM and stay above the minimum value of calcium-perchlorate almost the entire first half of the martian year. However during this time the temperature levels stay too low, only going above eutectic levels from the middle of the northern summer. By this time the relative humidity levels show a significant decrease and stay below 0.5 for the rest of the year. \\ 

\begin{figure}[H]
    \centering
    \includegraphics[width=\linewidth]{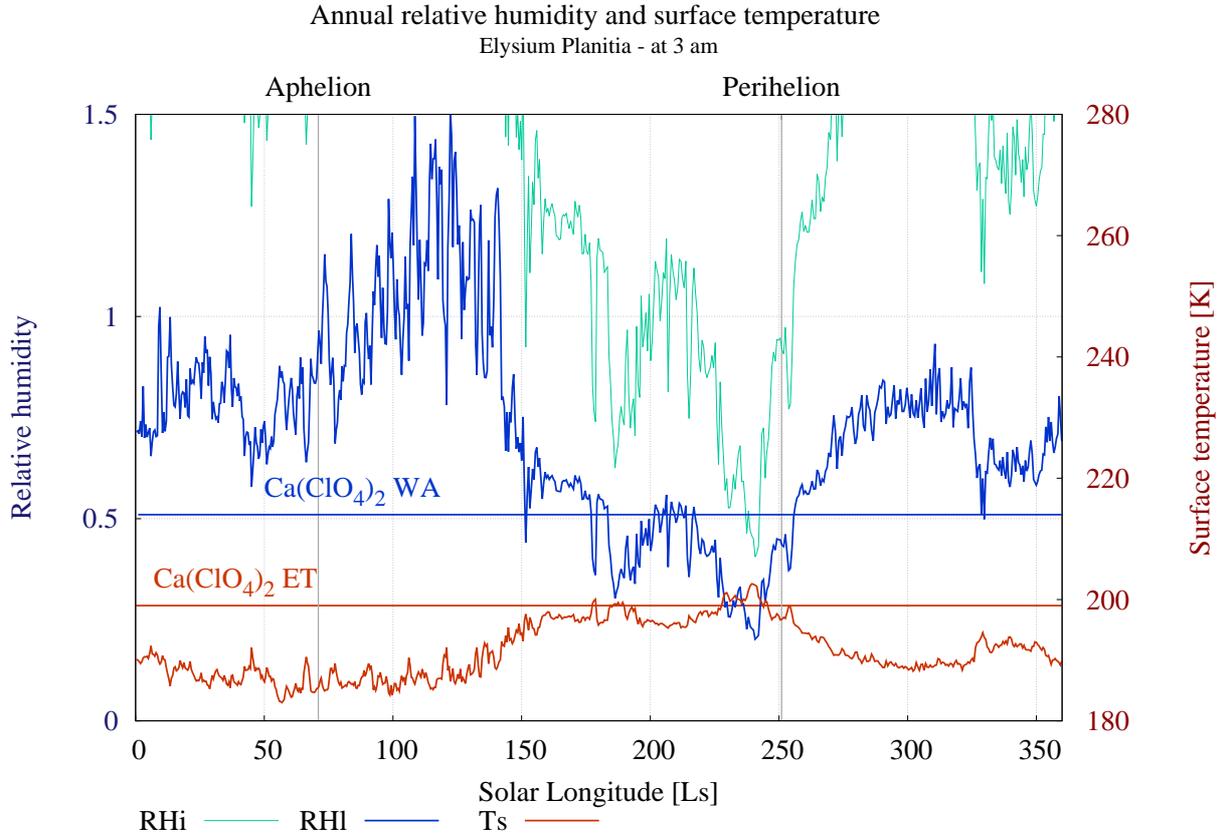}
    \caption{Annual variations of relative humidity and surface temperature at 3 AM local time at Elysium Planitia.}
    \label{fig:annual3am}
\end{figure}

Figure \ref{fig:annual3am} represents a non-ideal scenario. Here we can see that at 3 AM at Elysium Planitia the temperature stays below the minimum threshold value for the entire martian year except for a brief period around Ls 240$^{\circ}$. The relative humidity values peak at 3 AM and reach as far as 1.5 after the northern summer solstice, but dip under between Ls 150 $^{\circ}$ and 250 $^{\circ}$. This results in possibly no chance for liquefaction regardless of the briefly ideal temperature period. However, even if the surface temperatures would be high enough to reach the eutectic value between Ls 0$^{\circ}$ - 150$^{\circ}$ and at the end of the year, the relative humidity with respect to ice is elevated well beyond 1 meaning a more likely scenario of ice formation, than brines. \\

\subsection{Daily variations}
\label{subsec:daily}

This subsection contains the representative selection of daily curves investigated according to the prediction of the annual graphs. The cumulative results are depicted in Figure \ref{fig:allidrhts} shown in \ref{subsec:ideal}. Please note, that the daily graphs are centered around 3 AM and not 12 PM to position the potentially ideal nighttime in the middle of the diagram. \\

\begin{figure}[H]
    \centering
    \includegraphics[width=\linewidth]{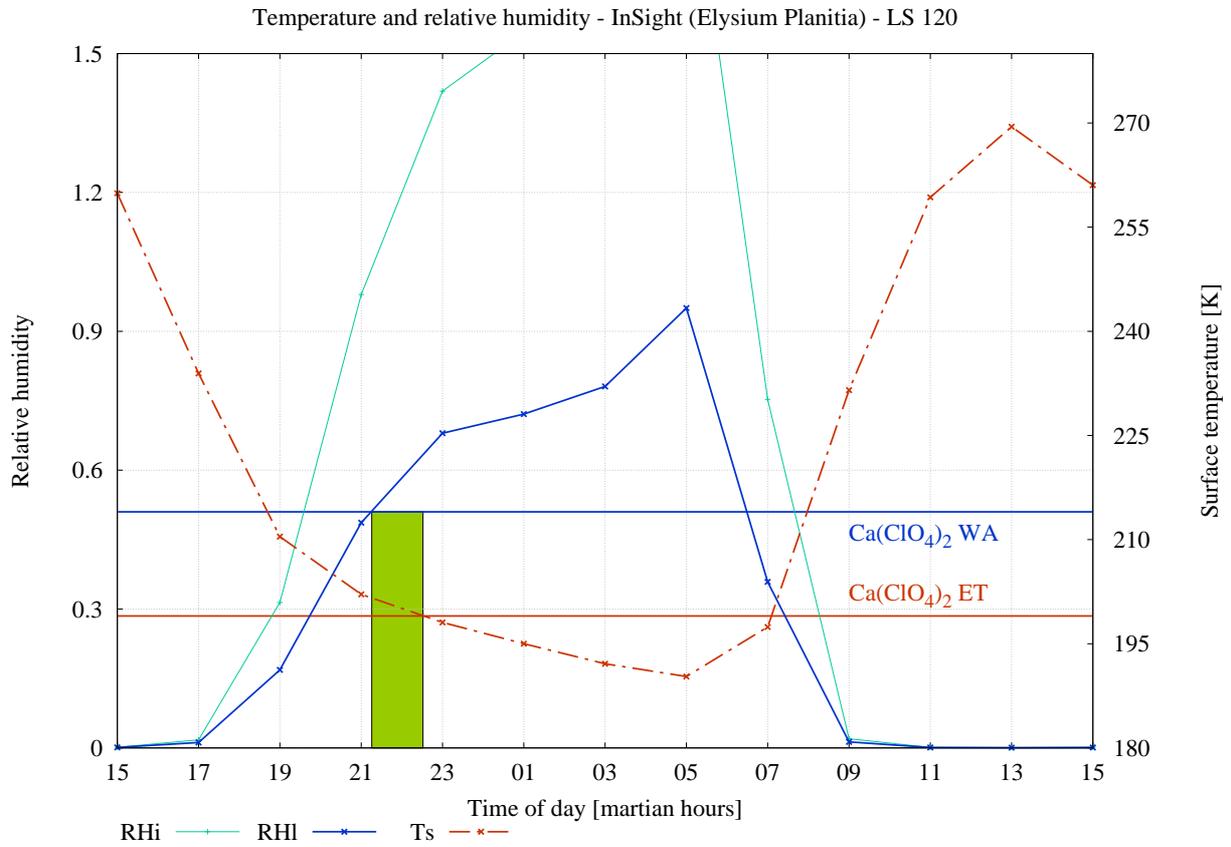}
    \caption{Daily variations of near-surface relative humidity and surface temperature at Ls 120$^\circ$. Here we can see a typical example of a possibly ideal period before midnight.}
    \label{fig:ls120idd}
\end{figure}

\begin{figure}[H]
   \centering
    \includegraphics[width=\linewidth]{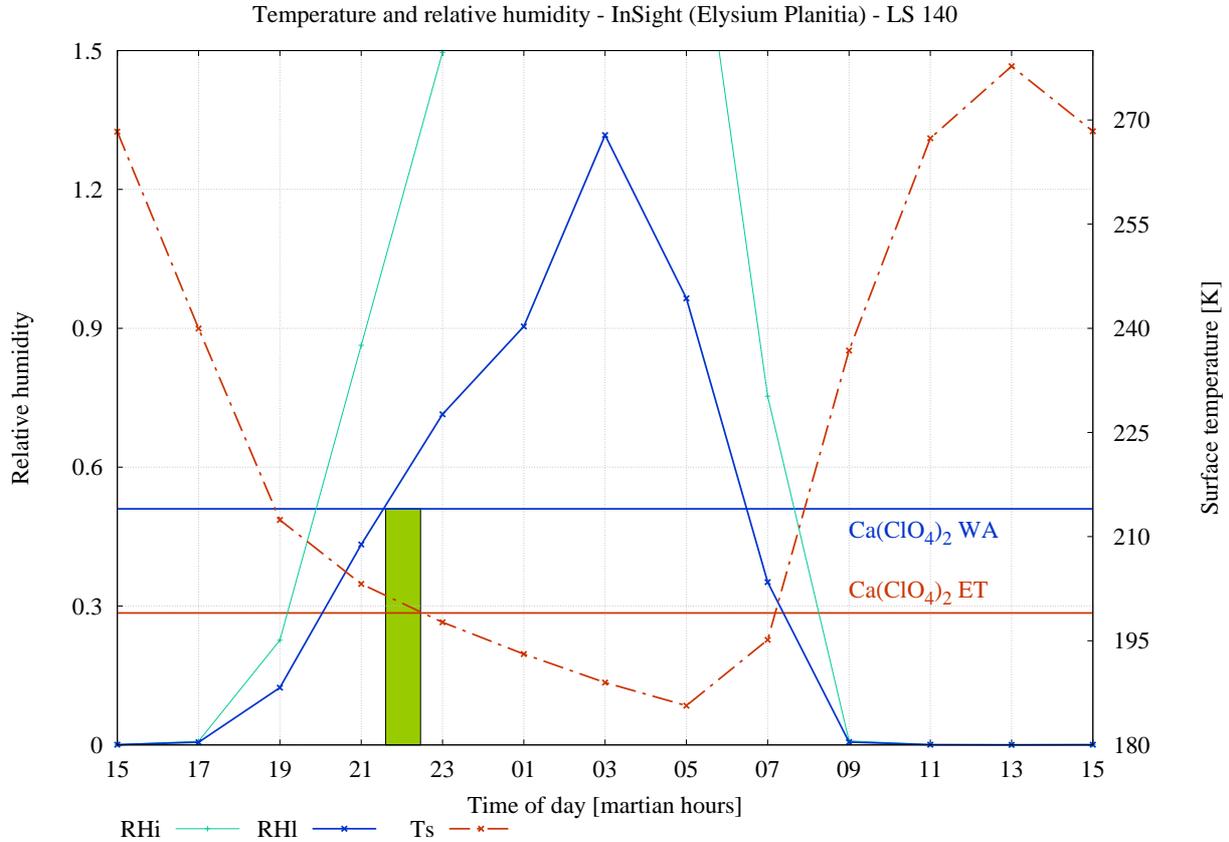}
    \caption{Daily variations of near-surface relative humidity and surface temperature at Ls 140$^\circ$. It shows that the approximately 1 hour period is shorter than at Ls 120$^{\circ}$ (Fig. \ref{fig:ls120idd}.})
   \label{fig:ls235idd}
\end{figure}

Figure \ref{fig:ls120idd} illustrates the most characteristic potentially ideal period distribution with an almost 2 hour long possible ideal period in the late evening. All solar longitude values investigated for daily variability show the same behaviour of an approximately 0.5-2 hour long window centered around 9 PM. As it was shown in the annual graph (Fig. \ref{fig:annual9pm}), the daily graphs shown a chance for deliquescence between Ls 90$^{\circ}$ - 140$^{\circ}$. Between these the possibly ideal time windows increase from as short as 30 minutes to almost 2 hours at Ls 120$^{\circ}$, then gradually get shorter after that and disappear after Ls 150$^{\circ}$. \\ 

\begin{figure}[H]
    \centering
    \includegraphics[width=\linewidth]{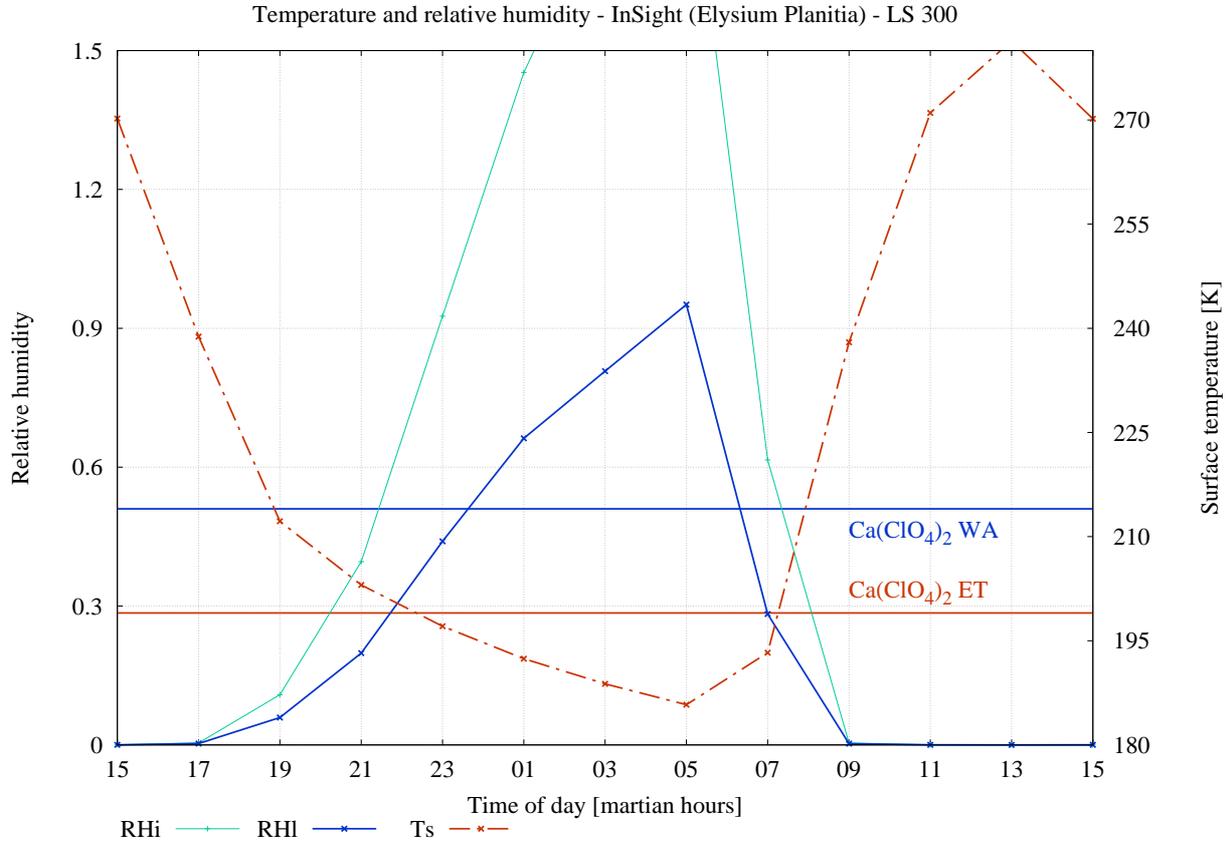}
    \caption{Daily variations of near-surface relative humidity and surface temperature at Ls 350$^\circ$. This figure shows a non-ideal scenario, where we can observe the typical problem of deliquescence with high enough temperatures and high enough relative humidities out of sync.}
    \label{fig:ls350d}
\end{figure}

Figure \ref{fig:ls350d} depicts a non-ideal day for deliquescence. This case illustrates well the difficulty of finding the optimal time periods with the temperatures just below the ET level of calcium-perchlorate when relative humidity is high enough and vice versa. \\

\subsection{Summary of ideal daily periods}
\label{subsec:ideal}

This subsection contains the two cumulative figures showing all the ideal days throughout one Martian year. The different local times are indicated by various colors and shapes. \\

\begin{figure}[H]
    \centering
    \begin{subfigure}[b]{0.49\textwidth}
    \includegraphics[width=\columnwidth]{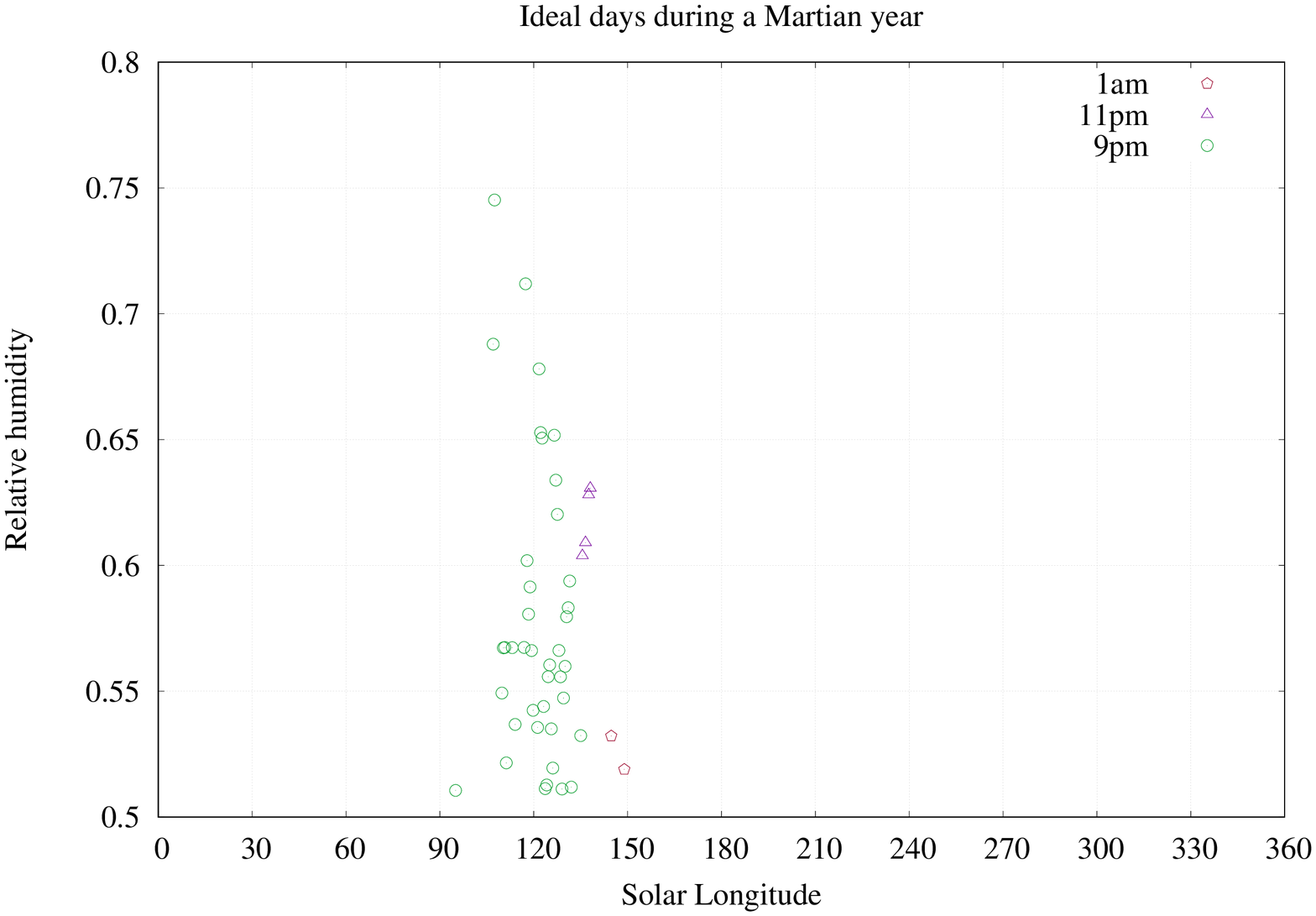}
    \end{subfigure}
    \begin{subfigure}[b]{0.49\textwidth}
    \includegraphics[width=\columnwidth]{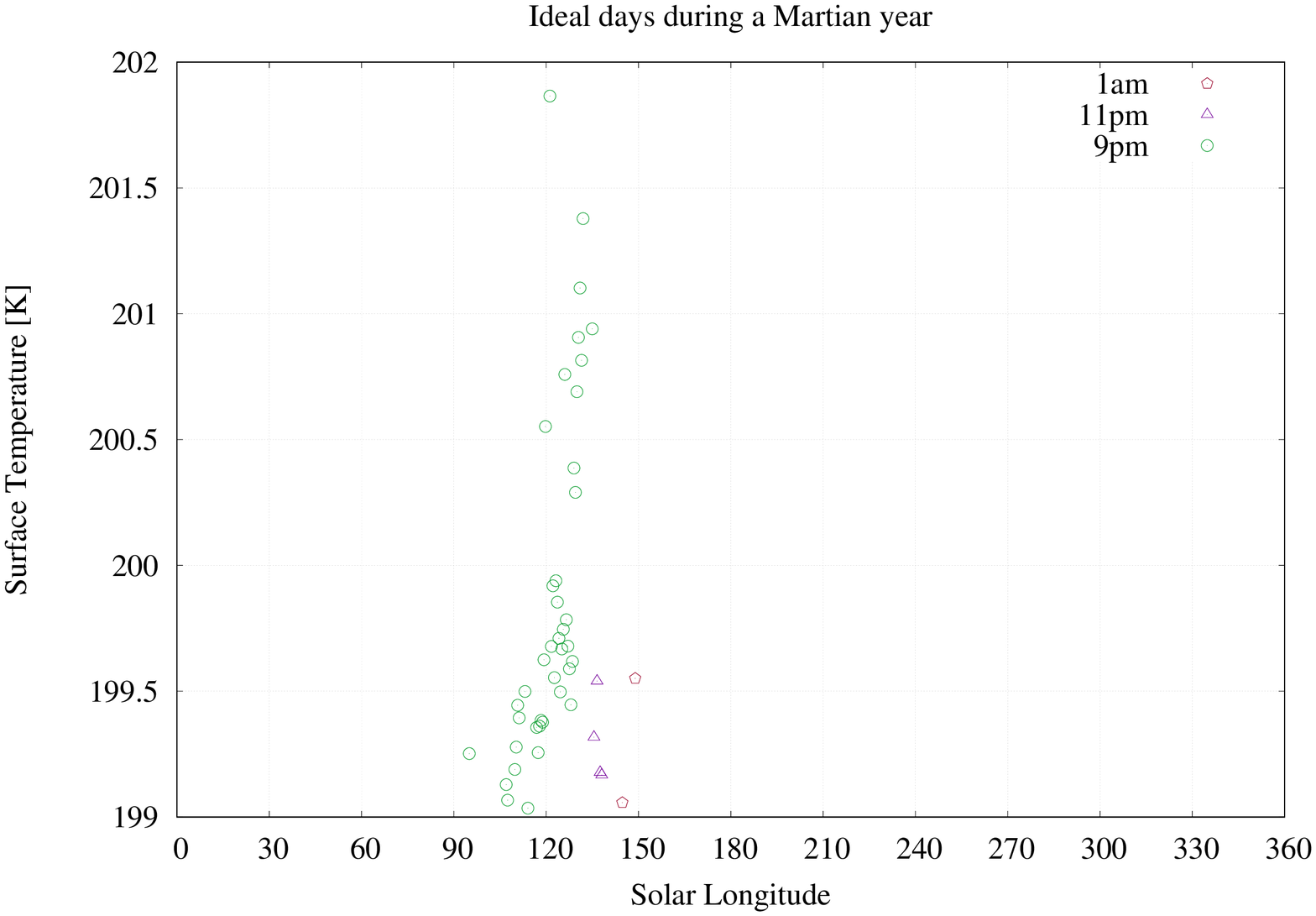}
    \end{subfigure}
    \caption{Summary of the ideal times throughout a Martian year illustrated with various colours and shapes for different local times. We can observe, that the possible window for deliquescence is concentrated between Ls 90$^{\circ}$ and Ls 150$^{\circ}$.}
    \label{fig:allidrhts}
\end{figure}

In Figure \ref{fig:allidrhts} we can see the ideal days in a year versus the relative humidity at the optimal time and in Figure \ref{fig:allidrhts} versus the surface temperature. The tendencies implied in the selection of annual curves in \ref{subsec:annual} show up nicely by the concentrated group centered around Ls 120$^{\circ}$. 9 PM local time dominates the group with the highest relative humidity values scattered in the narrow region around Ls 120$^{\circ}$, but note that there is no saturation. At 1 AM there is only four points indicating a very brief potential period around Ls 135$^{\circ}$. There is 2 points indicating possible deliquescence at 11 PM at the right end of the group just before Ls 150$^{\circ}$, where the relative humidity is just above the minimum value for calcium-perchlorate deliquescence. Showing the ideal days in relation to the surface temperature, the group appears even narrower with the temperatures peaking around Ls 130$^{\circ}$ at 9 PM. We can see that the temperature, while staying above the necessary eutectic value, never goes very high. \\

\section{Discussion}
\label{sec:discussion}

With the progress of Martian research related to liquid water, the investigation of near-surface conditions and relative humidity is getting more important, as understanding the possible ideal times for liquid water appearances could play a role in the planning of future Mars missions and in-situ measurement strategies. However, due to the complexity of processes close to the surface, the field of modelling the near-surface relative humidity research is still developing. Only two of all the spacecrafts that landed on Mars carried near-surface relative humidity sensors, the Phoenix and the Mars Science Laboratory \citep{zent2009,gomez2012}. Our research shows a method of investigating the possible ideal periods for deliquescence of hygroscopic salts at the landing site of the latest Mars mission, the InSight at Elysium Planitia. This can be used to examine some conditions of interest in future studies, and point to aspects, which could be the target of detailed analysis using the meteorological data from InSight to improve the estimation for the deliquescence process. \\

In this study we model the near-surface relative humidity and surface temperature values at the landing site of InSight (4.5$^\circ$N 135.9$^\circ$E). Through studying their annual variations as well as the daily cycles, we identify possible ideal times for deliquescence of Mars relevant hygroscopic salts, from which calcium-perchlorate proved to be the most likely candidate. We find, that ideal time windows could occur for a well defined time window between Ls 90$^{\circ}$ - 150$^{\circ}$ between 9 PM and 1 AM. The daily results show no such instances when the presumed ideal times could be continuous through the night, the possible ideal periods occur before midnight for approximately 0.5 - 2 hours centered around 9 PM.  Our findings of relative humidity values and daily cycle are consistent with the recalibrated Phoenix RH sensor data shown in \citet{fischer2018}. The daily course of relative humidity and temperature changes are in good agreement with previous measurements \citep{savi2019,riveravalentin2015,savijarvi2015}, while the annual cycle is consistent with the daily maximum RH measurements of the first 1258 sols of the MSL mission (MY 31-33) \citep{martinez2017}. The highest annual values occur around the early northern summer (Ls 90$^\circ$ - Ls 120$^\circ$). Our assumptions fit well in the search for hydration, for example the research of \citet{farris2018}, who argues that adsorption, coupled with the cohesive nature of the regolith could serve as an active water vapor sink in agreement with \citep{smith2009,riveravalentin2015,arvidson2009}. However \citet{gough2019} showed that while hydrated Ca(ClO$_4$)$_2$ or potentially a brine could exist, it is not likely to experience dehydration/hydration cycles via diurnal interaction with atmospheric water vapor. In their work they found that Ca(ClO$_4$)$_2$ will not dehydrate back to the anhydrous phase until the temperatures reach 298 K or warmer, and even in the warmest time of the day at Elysium Planitia it never went above 290 K according to our calculations, which is important to take into consideration. \\

The annual variability of surface temperatures and relative humidity shows a seasonal change, and RH typically increases towards perihelion, however the nighttime surface temperatures are consistent within a degree or two during the aphelion season. Comparing the model results with the globally averaged nighttime surface temperatures measured by TES \citep{smith2004}, they are coherent as our results also show an increase from approximately Ls 120$^\circ$ and decrease from approximately Ls 150$^\circ$. Looking at the diurnal curves, the temperature steadily decreases and dips below the eutectic value around the middle of the night in all instances. \\

The possibility of deliquescence has been investigated previously at former landing sites and the potential landing sites of the ExoMars rover in \citet{pal2017}. The possibly not ideal shorter period in the middle of the night was also visible from the modelling results at Aeolis Palus (137.44$^\circ$E, 4.59$^\circ$S), the landing site of Curiosity, and at Oxia Planum (333.5$^\circ$E, 18.2$^\circ$N). Both of these locations, the gap in the possible ideal period was caused by the temperature just dipping below the eutectic levels. It is important to note that in those cases the relative humidity was calculated with respect to ice and not liquid. The precise measurements of InSight could decide which daily behaviour is right, whether the whole night is ideal or there is a much longer pause due to the temperatures decreasing more quickly or if only the early night hours could be ideal as shown above. It could also shed light on the validity of our annual calculations as the modeled temperature values stay close to the eutectic value. With temperature data from InSight, we could gain more information on the actual behaviour of the climate near the surface during the night. \\

In our study we assume water vapor volume mixing ratio to be well mixed between the surface and approximately 4 m above the surface. Thus while calculating the relative humidity values we compare atmospheric water vapor with surface temperature, similar to the earlier work of \citet{pal2017, pal2019}. While this approach does not result in completely accurate data due to the complexity of near-surface processes, it gives reasonable and best among the currently available estimates. In our daily graphs we collected data at every 2 hours from the GCM model. Consequently the time windows we marked as possibly ideal should be seen as a good approximation rather than fixed local times. \\

Our results suggest that future in-situ measurements searching for thin liquid films at near equatorial region by deliquescence should focus on the early night hours from the beginning of the northern summer until late summer. As the RH values decrease towards the northern autumn and perihelion, the early night can be no longer ideal, even though the temperatures are over the eutectic value. Although our results show that there is only a concentrated time window between appoximately Ls 90$^{\circ}$ - 150$^{\circ}$, this could be the consequence of the limitations of our study detailed above, and needs further research or possible clarification with the help of the InSight measurements. Our method of calculations could also be expanded to other areas of interest in future works. \citet{gough2014} showed that multicomponent brines would have lower DRH levels and may be more stable on the Martian surface, thus could be a good next step to investigate. A way of estimating the amount of liquid water forming could also be an interesting challenge for upcoming studies. \\

The results show that mostly around 9 PM could be ideal for deliquescence, but the presumed periods are not continuous due to temperature fluctuating near the eutectic value. The precise measurements from InSight could be used to compare with the model results to determine if the periods are continuous or not or even if they start sooner from around Ls 75$^{\circ}$ when the calculated relative humidity first reaches above the minimum levels. More accurate in-situ temperature data could be useful at the 11 PM data as well, where the beginning of the year, between Ls 0 - 150$^\circ$ the temperature stays close to the threshold value. It would be also interesting to see, that at 3 AM between Ls 150 - 250$^\circ$ the temperature indeed stays just below ET level or not. The daily graphs show a typical behaviour of temperature dropping below ET value around midnight. InSight in-situ ground truth measurements could shed light on if the decrease in temperature is valid or the potentially ideal periods could last for the whole duration of the night as the relative humidity values stay above minimum levels until the early morning hours. \\

\section{Conclusion}
\label{sec:conclusion}

In our study we calculated near-surface relative humidity values from the output of LMDZ GCM climate model of Mars and compared them with surface temperature data to find possible ideal time periods for the deliquescence of certain hygroscopic salts at the landing site of InSight. We investigated the annual variations to search ideal times throughout the Martian year. After the annual study we examined daily cycles and presumed ideal time windows according to local times. \\ 

We have found that at Elysium Planitia, the landing site of InSight potentially ideal time periods for deliquescence could occur at a concentrated period between approximately Ls 90$^{\circ}$ - 150$^{\circ}$. Our daily cycle studies show that the ideal local times fluctuate around 9 PM showing 0.5 - 2 hours of possible deliquescence before midnight. In all instances we found no such day where the ideal period could last all night long due to the temperature dropping below the necessary value. The precise temperature measurements of InSight as in-situ measured ground truth could help determine if the circumstances are ideal continuously or the ideal periods truly end before midnight even with the relative humidity staying elevated until the early morning hours.\\

Our results suggest time periods which could be interesting in the search for liquid brine appearance throughout a Martian year. We also propose a method to look for the possibly ideal times from model calculations as with the HABIT equipment on the upcoming ExoMars 2020 lander deliquescence studies require more and more attention, as ExoMars will be also situated close to the equator. \\

\section{Acknowledgement}
\label{sec:ack}

This work was supported by the EXODRILTECH project of ESA and the Excellence of Strategic R\&D centres (GINOP-2.3.2-15-2016-00003) project of NKFIH and the related H2020 fund, the COOP-NN-116927 project of NKFIH and the TD1308 \textit{Origins and evolution of life on Earth and in the Universe} COST actions number 39045 and 39078. The NetCDF files were visualized with the NASA GISS Panoply viewer developed by Dr. Robert B. Schmunk. We also express our gratitude to our anonymous reviewer, who greatly improved our manuscript and the validity of our work. \\





\bibliographystyle{elsarticle-harv}

\bibliography{ref,marsref}

\begin{thebibliography}{65}
\expandafter\ifx\csname natexlab\endcsname\relax\def\natexlab#1{#1}\fi
\expandafter\ifx\csname url\endcsname\relax
  \def\url#1{\texttt{#1}}\fi
\expandafter\ifx\csname urlprefix\endcsname\relax\def\urlprefix{URL }\fi

\bibitem[{{Arvidson} et~al.(2009){Arvidson}, {Bonitz}, {Robinson}, {Carsten},
  {Volpe}, {Trebi-Ollennu}, {Mellon}, {Chu}, {Davis}, {Wilson}, {Shaw},
  {Greenberger}, {Siebach}, {Stein}, {Cull}, {Goetz}, {Morris}, {Ming},
  {Keller}, {Lemmon}, {Sizemore}, and {Mehta}}]{arvidson2009}
{Arvidson}, R.~E., {Bonitz}, R.~G., {Robinson}, M.~L., {Carsten}, J.~L.,
  {Volpe}, R.~A., {Trebi-Ollennu}, A., {Mellon}, M.~T., {Chu}, P.~C., {Davis},
  K.~R., {Wilson}, J.~J., {Shaw}, A.~S., {Greenberger}, R.~N., {Siebach},
  K.~L., {Stein}, T.~C., {Cull}, S.~C., {Goetz}, W., {Morris}, R.~V., {Ming},
  D.~W., {Keller}, H.~U., {Lemmon}, M.~T., {Sizemore}, H.~G., {Mehta}, M., Oct
  2009. {Results from the Mars Phoenix Lander Robotic Arm experiment}. Journal
  of Geophysical Research (Planets) 114, E00E02.

\bibitem[{{Atreya} et~al.(2006){Atreya}, {Wong}, {Renno}, {Farrell}, {Delory},
  {Sentman}, {Cummer}, {Marshall}, {Rafkin}, and {Catling}}]{atreya2006}
{Atreya}, S.~K., {Wong}, A.-S., {Renno}, N.~O., {Farrell}, W.~M., {Delory},
  G.~T., {Sentman}, D.~D., {Cummer}, S.~A., {Marshall}, J.~R., {Rafkin},
  S.~C.~R., {Catling}, D.~C., Jun. 2006. {Oxidant Enhancement in Martian Dust
  Devils and Storms: Implications for Life and Habitability}. Astrobiology 6,
  439--450.

\bibitem[{{Banfield} and {InSight Science Team}(2014)}]{banfield2014}
{Banfield}, D., {InSight Science Team}, Jan 2014. {Atmospheric Observations
  from the Mars Insight Mission}. In: {Forget}, F., {Millour}, M. (Eds.), Mars
  Atmosphere: Modelling and Observation, 5th International Workshop. p. 4304.

\bibitem[{{Berger} et~al.(2016){Berger}, {Schmidt}, {Gellert}, {Campbell},
  {King}, {Flemming}, {Ming}, {Clark}, {Pradler}, {VanBommel}, {Minitti},
  {Fair{\'e}n}, {Boyd}, {Thompson}, {Perrett}, {Elliott}, and
  {Desouza}}]{berger2016}
{Berger}, J.~A., {Schmidt}, M.~E., {Gellert}, R., {Campbell}, J.~L., {King},
  P.~L., {Flemming}, R.~L., {Ming}, D.~W., {Clark}, B.~C., {Pradler}, I.,
  {VanBommel}, S.~J.~V., {Minitti}, M.~E., {Fair{\'e}n}, A.~G., {Boyd}, N.~I.,
  {Thompson}, L.~M., {Perrett}, G.~M., {Elliott}, B.~E., {Desouza}, E., Jan.
  2016. {A global Mars dust composition refined by the Alpha-Particle X-ray
  Spectrometer in Gale Crater}. Geophysics Research Letters 43, 67--75.

\bibitem[{{Brass}(1980)}]{brass1980}
{Brass}, G.~W., Apr. 1980. {Stability of brines on Mars}. Icarus 42, 20--28.

\bibitem[{{Chevrier} et~al.(2009){Chevrier}, {Hanley}, and
  {Altheide}}]{chevrier2009}
{Chevrier}, V.~F., {Hanley}, J., {Altheide}, T.~S., May 2009. {Stability of
  perchlorate hydrates and their liquid solutions at the Phoenix landing site,
  Mars}. Geophys. Res. Lett. 36, L10202.

\bibitem[{{Clancy} et~al.(2000){Clancy}, {Sandor}, {Wolff}, {Christensen},
  {Smith}, {Pearl}, {Conrath}, and {Wilson}}]{clancy2000}
{Clancy}, R.~T., {Sandor}, B.~J., {Wolff}, M.~J., {Christensen}, P.~R.,
  {Smith}, M.~D., {Pearl}, J.~C., {Conrath}, B.~J., {Wilson}, R.~J., Apr. 2000.
  {An intercomparison of ground-based millimeter, MGS TES, and Viking
  atmospheric temperature measurements: Seasonal and interannual variability of
  temperatures and dust loading in the global Mars atmosphere}. Journal of
  Geophysical Research 105, 9553--9572.

\bibitem[{{Clark} and {Kounaves}(2016)}]{clark2016}
{Clark}, B.~C., {Kounaves}, S.~P., Oct. 2016. {Evidence for the distribution of
  perchlorates on Mars}. International Journal of Astrobiology 15, 311--318.

\bibitem[{{Clow}(1987)}]{clow1987}
{Clow}, G.~D., Oct. 1987. {Generation of liquid water on Mars through the
  melting of a dusty snowpack}. Icarus 72, 95--127.

\bibitem[{{Davila} et~al.(2010){Davila}, {Duport}, {Melchiorri}, {J{\"a}nchen},
  {Valea}, {de los Rios}, {Fair{\'e}n}, {M{\"o}hlmann}, {McKay}, {Ascaso}, and
  {Wierzchos}}]{davila2010}
{Davila}, A.~F., {Duport}, L.~G., {Melchiorri}, R., {J{\"a}nchen}, J., {Valea},
  S., {de los Rios}, A., {Fair{\'e}n}, A.~G., {M{\"o}hlmann}, D., {McKay},
  C.~P., {Ascaso}, C., {Wierzchos}, J., Aug. 2010. {Hygroscopic Salts and the
  Potential for Life on Mars}. Astrobiology 10, 617--628.

\bibitem[{{Dundas}(2018)}]{dundas2018}
{Dundas}, C.~M., Mar 2018. {HiRISE Observations of New Martian Slope Streaks}.
  In: Lunar and Planetary Science Conference. p. 2026.

\bibitem[{{Elsenousy} et~al.(2015){Elsenousy}, {Hanley}, and
  {Chevrier}}]{elsenousy2015}
{Elsenousy}, A., {Hanley}, J., {Chevrier}, V.~F., Jul. 2015. {Effect of
  evaporation and freezing on the salt paragenesis and habitability of brines
  at the Phoenix landing site}. Earth and Planetary Science Letters 421,
  39--46.

\bibitem[{{Farris} et~al.(2018){Farris}, {Conner}, {Chevrier}, and
  {Rivera-Valentin}}]{farris2018}
{Farris}, H.~N., {Conner}, M.~B., {Chevrier}, V.~F., {Rivera-Valentin}, E.~G.,
  Jul 2018. {Adsorption driven regolith-atmospheric water vapor transfer on
  Mars: An analysis of Phoenix TECP data}. Icarus 308, 71--75.

\bibitem[{{Fischer} et~al.(2018){Fischer}, {Martinez}, and
  {Renno}}]{fischer2018}
{Fischer}, E., {Martinez}, G., {Renno}, N., Sep 2018. {Analysis of Recalibrated
  Phoenix Relative Humidity Sensor Data}. In: European Planetary Science
  Congress. pp. EPSC2018--564.

\bibitem[{{Forget} et~al.(1999){Forget}, {Hourdin}, {Fournier}, {Hourdin},
  {Talagrand}, {Collins}, {Lewis}, {Read}, and {Huot}}]{forget1999}
{Forget}, F., {Hourdin}, F., {Fournier}, R., {Hourdin}, C., {Talagrand}, O.,
  {Collins}, M., {Lewis}, S.~R., {Read}, P.~L., {Huot}, J.-P., Oct. 1999.
  {Improved general circulation models of the Martian atmosphere from the
  surface to above 80 km}. J. Geophys. Res. 104, 24155--24176.

\bibitem[{{Goff} and {Gratch}(1946)}]{goff1946}
{Goff}, J.~A., {Gratch}, S., 1946. {Low pressure properties of water from -160
  to 212 F}. Trans. Amer. Soc. Heat. Vent. Eng. 52, 95.

\bibitem[{{Golombek} et~al.(2017){Golombek}, {Kipp}, {Warner}, {Daubar},
  {Fergason}, {Kirk}, {Beyer}, {Huertas}, {Piqueux}, {Putzig}, {Campbell},
  {Morgan}, {Charalambous}, {Pike}, {Gwinner}, {Calef}, {Kass}, {Mischna},
  {Ashley}, {Bloom}, {Wigton}, {Hare}, {Schwartz}, {Gengl}, {Redmond},
  {Trautman}, {Sweeney}, {Grima}, {Smith}, {Sklyanskiy}, {Lisano}, {Benardini},
  {Smrekar}, {Lognonn{\'e}}, and {Banerdt}}]{golombek2017}
{Golombek}, M., {Kipp}, D., {Warner}, N., {Daubar}, I.~J., {Fergason}, R.,
  {Kirk}, R.~L., {Beyer}, R., {Huertas}, A., {Piqueux}, S., {Putzig}, N.~E.,
  {Campbell}, B.~A., {Morgan}, G.~A., {Charalambous}, C., {Pike}, W.~T.,
  {Gwinner}, K., {Calef}, F., {Kass}, D., {Mischna}, M., {Ashley}, J., {Bloom},
  C., {Wigton}, N., {Hare}, T., {Schwartz}, C., {Gengl}, H., {Redmond}, L.,
  {Trautman}, M., {Sweeney}, J., {Grima}, C., {Smith}, I.~B., {Sklyanskiy}, E.,
  {Lisano}, M., {Benardini}, J., {Smrekar}, S., {Lognonn{\'e}}, P., {Banerdt},
  W.~B., Oct. 2017. {Selection of the InSight Landing Site}. Space Sci. Rev.
  211, 5--95.

\bibitem[{{Golombek} et~al.(2016){Golombek}, {Warner}, {Daubar}, {Kipp},
  {Huertas}, {Beyer}, {Piqueux}, {Putzig}, {Calef}, and
  {Banerdt}}]{golombek2016}
{Golombek}, M., {Warner}, N., {Daubar}, I.~J., {Kipp}, D., {Huertas}, A.,
  {Beyer}, R., {Piqueux}, S., {Putzig}, N.~E., {Calef}, F., {Banerdt}, W.~B.,
  Mar. 2016. {Surface and Subsurface Characteristics of Western Elysium
  Planitia, Mars}. In: Lunar and Planetary Science Conference. Vol.~47 of Lunar
  and Planetary Science Conference. p. 1572.

\bibitem[{{G{\'o}mez-Elvira} et~al.(2012){G{\'o}mez-Elvira}, {Armiens},
  {Casta{\~n}er}, {Dom{\'\i}nguez}, {Genzer}, {G{\'o}mez}, {Haberle}, {Harri},
  {Jim{\'e}nez}, and {Kahanp{\"a}{\"a}}}]{gomez2012}
{G{\'o}mez-Elvira}, J., {Armiens}, C., {Casta{\~n}er}, L., {Dom{\'\i}nguez},
  M., {Genzer}, M., {G{\'o}mez}, F., {Haberle}, R., {Harri}, A.~M.,
  {Jim{\'e}nez}, V., {Kahanp{\"a}{\"a}}, H., Sep 2012. {REMS: The Environmental
  Sensor Suite for the Mars Science Laboratory Rover}. Space Science Reviews
  170~(1-4), 583--640.

\bibitem[{{Gough} et~al.(2011){Gough}, {Chevrier}, {Baustian}, {Wise}, and
  {Tolbert}}]{gough2011}
{Gough}, R.~V., {Chevrier}, V.~F., {Baustian}, K.~J., {Wise}, M.~E., {Tolbert},
  M.~A., Dec. 2011. {Laboratory studies of perchlorate phase transitions:
  Support for metastable aqueous perchlorate solutions on Mars}. Earth and
  Planetary Science Letters 312, 371--377.

\bibitem[{{Gough} et~al.(2014){Gough}, {Chevrier}, and {Tolbert}}]{gough2014}
{Gough}, R.~V., {Chevrier}, V.~F., {Tolbert}, M.~A., May 2014. {Formation of
  aqueous solutions on Mars via deliquescence of chloride-perchlorate binary
  mixtures}. Earth and Planetary Science Letters 393, 73--82.

\bibitem[{{Gough} et~al.(2019){Gough}, {Primm}, {Rivera-Valent{\'\i}n},
  {Mart{\'\i}nez}, and {Tolbert}}]{gough2019}
{Gough}, R.~V., {Primm}, K.~M., {Rivera-Valent{\'\i}n}, E.~G., {Mart{\'\i}nez},
  G.~M., {Tolbert}, M.~A., Mar 2019. {Solid-solid hydration and dehydration of
  Mars-relevant chlorine salts: Implications for Gale Crater and RSL
  locations}. Icarus 321, 1--13.

\bibitem[{{Haberle} et~al.(2001){Haberle}, {McKay}, {Schaeffer}, {Cabrol},
  {Grin}, {Zent}, and {Quinn}}]{haberle2001}
{Haberle}, R.~M., {McKay}, C.~P., {Schaeffer}, J., {Cabrol}, N.~A., {Grin},
  E.~A., {Zent}, A.~P., {Quinn}, R., Oct. 2001. {On the possibility of liquid
  water on present-day Mars}. J. Geophys. Res. 106, 23317--23326.

\bibitem[{{Hecht}(2002)}]{hecht2002}
{Hecht}, M.~H., Apr. 2002. {Metastability of Liquid Water on Mars}. Icarus 156,
  373--386.

\bibitem[{{Hecht} et~al.(2009){Hecht}, {Catling}, {Clark}, {Deflores},
  {Gospodinova}, {Kapit}, {Kounaves}, {Ming}, {Quinn}, {West}, and
  {Young}}]{hecht2009}
{Hecht}, M.~H., {Catling}, D.~C., {Clark}, B.~C., {Deflores}, L.,
  {Gospodinova}, K., {Kapit}, J., {Kounaves}, S.~P., {Ming}, D.~W., {Quinn},
  R.~C., {West}, S.~J., {Young}, S.~M.~M., Mar. 2009. {Perchlorate in Martian
  Soil: Evidence and Implications}. In: Lunar and Planetary Science Conference.
  Vol.~40 of Lunar and Planetary Science Conference. p. 2420.

\bibitem[{{Kereszturi} et~al.(2009){Kereszturi}, {M{\"o}hlmann}, {Berczi},
  {Ganti}, {Kuti}, {Sik}, and {Horvath}}]{kereszturi2009}
{Kereszturi}, A., {M{\"o}hlmann}, D., {Berczi}, S., {Ganti}, T., {Kuti}, A.,
  {Sik}, A., {Horvath}, A., Jun. 2009. {Recent rheologic processes on dark
  polar dunes of Mars: Driven by interfacial water?} Icarus 201, 492--503.

\bibitem[{{Knauth} and {Burt}(2002)}]{knauth2002}
{Knauth}, L.~P., {Burt}, D.~M., Jul. 2002. {Eutectic Brines on Mars: Origin and
  Possible Relation to Young Seepage Features}. Icarus 158, 267--271.

\bibitem[{{Kossacki} and {Markiewicz}(2004)}]{kossacki2004}
{Kossacki}, K.~J., {Markiewicz}, W.~J., Oct. 2004. {Seasonal melting of surface
  water ice condensing in martian gullies}. Icarus 171, 272--283.

\bibitem[{Kossacki(2008)}]{kossacki2008}
Kossacki, K.J., M.~W., 2008. Martian hill-gullies, surface and sub-surface
  moisture. In: In: Mars Water Cycle Workshop, Paris.

\bibitem[{{List}(1984)}]{list1984}
{List}, R.~J., 1984. {Smithsonian Meteorological Tables}. 6th ed. Smithsonian
  Institution Press, 347--350.

\bibitem[{Martin-Torres et~al.(2015)Martin-Torres, Zorzano, Valentin-Serrano,
  Harri, Genzer, Kemppinen, Rivera-Valentin, Jun, Wray, Bo~Madsen, Goetz,
  McEwen, Hardgrove, Renno, Chevrier, Mischna, Navarro-Gonzalez,
  Martinez-Frias, Conrad, McConnochie, Cockell, Berger, R.~Vasavada, Sumner,
  and Vaniman}]{torres2015}
Martin-Torres, F.~J., Zorzano, M.-P., Valentin-Serrano, P., Harri, A.-M.,
  Genzer, M., Kemppinen, O., Rivera-Valentin, E.~G., Jun, I., Wray, J.,
  Bo~Madsen, M., Goetz, W., McEwen, A.~S., Hardgrove, C., Renno, N., Chevrier,
  V.~F., Mischna, M., Navarro-Gonzalez, R., Martinez-Frias, J., Conrad, P.,
  McConnochie, T., Cockell, C., Berger, G., R.~Vasavada, A., Sumner, D.,
  Vaniman, D., May 2015. Transient liquid water and water activity at gale
  crater on mars. Nature Geoscience 8~(5), 357--361.
\newline\urlprefix\url{http://dx.doi.org/10.1038/ngeo2412}

\bibitem[{{Mart{\'\i}nez} et~al.(2017){Mart{\'\i}nez}, {Newman}, {De
  Vicente-Retortillo}, {Fischer}, {Renno}, {Richardson}, {Fair{\'e}n},
  {Genzer}, {Guzewich}, and {Haberle}}]{martinez2017}
{Mart{\'\i}nez}, G.~M., {Newman}, C.~N., {De Vicente-Retortillo}, A.,
  {Fischer}, E., {Renno}, N.~O., {Richardson}, M.~I., {Fair{\'e}n}, A.~G.,
  {Genzer}, M., {Guzewich}, S.~D., {Haberle}, R.~M., Oct 2017. {The Modern
  Near-Surface Martian Climate: A Review of In-situ Meteorological Data from
  Viking to Curiosity}. Space Science Reviews 212~(1-2), 295--338.

\bibitem[{{Mart{\'{\i}}nez} and {Renno}(2013)}]{martinez2013}
{Mart{\'{\i}}nez}, G.~M., {Renno}, N.~O., Jun. 2013. {Water and Brines on Mars:
  Current Evidence and Implications for MSL}. Space Science Reviews 175,
  29--51.

\bibitem[{McEwen et~al.(2014)McEwen, Dundas, Mattson, Toigo, Ojha, Wray,
  Chojnacki, Byrne, Murchie, and Thomas}]{mcewen2014}
McEwen, A., Dundas, C., Mattson, S., Toigo, A., Ojha, L., Wray, J., Chojnacki,
  M., Byrne, S., Murchie, S., Thomas, N., 2014. Recurring slope lineae in
  equatorial regions of mars. Nature Geoscience 7~(1), 53--58.

\bibitem[{{McEwen} et~al.(2011){McEwen}, {Ojha}, {Dundas}, {Mattson}, {Byrne},
  {Wray}, {Cull}, {Murchie}, {Thomas}, and {Gulick}}]{mcewen2011}
{McEwen}, A.~S., {Ojha}, L., {Dundas}, C.~M., {Mattson}, S.~S., {Byrne}, S.,
  {Wray}, J.~J., {Cull}, S.~C., {Murchie}, S.~L., {Thomas}, N., {Gulick},
  V.~C., Aug. 2011. {Seasonal Flows on Warm Martian Slopes}. Science 333, 740.

\bibitem[{{McSween} and {Keil}(2000)}]{mcsween2000}
{McSween}, H.~Y., {Keil}, K., Jun. 2000. {Mixing relationships in the Martian
  regolith and the composition of globally homogeneous dust}. Geochimica
  Cosmochimica Acta 64, 2155--2166.

\bibitem[{{Mellon} and {Phillips}(2001)}]{Mellon2001}
{Mellon}, M.~T., {Phillips}, R.~J., Oct. 2001. {Recent gullies on Mars and the
  source of liquid water}. J. Geophys. Res. 106, 23165--23180.

\bibitem[{{Millour} et~al.(2014){Millour}, {Forget}, {Spiga}, {Navarro},
  {Madeleine}, {Montabone}, {Lefevre}, {Chaufray}, {Lopez-Valverde},
  {Gonzalez-Galindo}, {Lewis}, {Read}, {Desjean}, {Huot}, and {MCD/GCM
  Development Team}}]{millour2014}
{Millour}, E., {Forget}, F., {Spiga}, A., {Navarro}, T., {Madeleine}, J.~B.,
  {Montabone}, L., {Lefevre}, F., {Chaufray}, J.~Y., {Lopez-Valverde}, M.~A.,
  {Gonzalez-Galindo}, F., {Lewis}, S.~R., {Read}, P.~L., {Desjean}, M.~C.,
  {Huot}, J.~P., {MCD/GCM Development Team}, Jul 2014. {The Mars Climate
  Database (MCD version 5.1)}. In: Eighth International Conference on Mars.
  Vol. 1791. p. 1184.

\bibitem[{{Millour} et~al.(2015){Millour}, {Forget}, {Spiga}, {Navarro},
  {Madeleine}, {Montabone}, {Pottier}, {Lefevre}, {Montmessin}, {Chaufray},
  {Lopez-Valverde}, {Gonzalez-Galindo}, {Lewis}, {Read}, {Huot}, {Desjean}, and
  {MCD/GCM development Team}}]{millour2015}
{Millour}, E., {Forget}, F., {Spiga}, A., {Navarro}, T., {Madeleine}, J.~B.,
  {Montabone}, L., {Pottier}, A., {Lefevre}, F., {Montmessin}, F., {Chaufray},
  J.~Y., {Lopez-Valverde}, M.~A., {Gonzalez-Galindo}, F., {Lewis}, S.~R.,
  {Read}, P.~L., {Huot}, J.~P., {Desjean}, M.~C., {MCD/GCM development Team},
  Oct 2015. {The Mars Climate Database (MCD version 5.2)}. In: European
  Planetary Science Congress. pp. EPSC2015--438.

\bibitem[{{M{\"o}hlmann} and {Thomsen}(2011)}]{mohlmann2011}
{M{\"o}hlmann}, D., {Thomsen}, K., Mar. 2011. {Properties of cryobrines on
  Mars}. Icarus 212, 123--130.

\bibitem[{{M{\"o}hlmann}(2004)}]{mohlmann2004}
{M{\"o}hlmann}, D.~T.~F., Apr. 2004. {Water in the upper martian surface at
  mid- and low-latitudes: presence, state, and consequences}. Icarus 168,
  318--323.

\bibitem[{{Motazedian}(2003)}]{motazedian2003}
{Motazedian}, T., Mar. 2003. {Currently Flowing Water on Mars}. In: {Mackwell},
  S., {Stansbery}, E. (Eds.), Lunar and Planetary Science Conference. Vol.~34
  of Lunar and Planetary Science Conference.

\bibitem[{{Nakamura} and {Anderson}(1979)}]{nakamura1979}
{Nakamura}, Y., {Anderson}, D.~L., Jun. 1979. {Martian wind activity detected
  by a seismometer at Viking lander 2 site}. Geophysics Research Letters 6,
  499--502.

\bibitem[{{Navarro} et~al.(2014){Navarro}, {Madeleine}, {Forget}, {Spiga},
  {Millour}, {Montmessin}, and {M{\"a}{\"a}tt{\"a}nen}}]{navarro2014}
{Navarro}, T., {Madeleine}, J.-B., {Forget}, F., {Spiga}, A., {Millour}, E.,
  {Montmessin}, F., {M{\"a}{\"a}tt{\"a}nen}, A., Jul. 2014. {Global climate
  modeling of the Martian water cycle with improved microphysics and
  radiatively active water ice clouds}. Journal of Geophysical Research
  (Planets) 119, 1479--1495.

\bibitem[{{Ojha} et~al.(2015){Ojha}, {Wilhelm}, {Murchie}, {McEwen}, {Wray},
  {Hanley}, {Mass{\'e}}, and {Chojnacki}}]{ojha2015}
{Ojha}, L., {Wilhelm}, M.~B., {Murchie}, S.~L., {McEwen}, A.~S., {Wray}, J.~J.,
  {Hanley}, J., {Mass{\'e}}, M., {Chojnacki}, M., Nov. 2015. {Spectral evidence
  for hydrated salts in recurring slope lineae on Mars}. Nature Geoscience 8,
  829--832.

\bibitem[{{P{\'a}l} and {Kereszturi}(2017)}]{pal2017}
{P{\'a}l}, B., {Kereszturi}, {\'A}., Jan. 2017. {Possibility of microscopic
  liquid water formation at landing sites on Mars and their observational
  potential}. Icarus 282, 84--92.

\bibitem[{{Pan} and {Quantin}(2018)}]{pan2018}
{Pan}, L., {Quantin}, C., Mar. 2018. {Regional Geological Context of the
  InSight Landing Site from Mineralogy and Stratigraphy}. In: Lunar and
  Planetary Science Conference. Vol.~49 of Lunar and Planetary Science
  Conference. p. 1918.

\bibitem[{Pál et~al.(2019)Pál, Ákos Kereszturi, Forget, and Smith}]{pal2019}
Pál, B., Ákos Kereszturi, Forget, F., Smith, M.~D., 2019. Global seasonal
  variations of the near-surface relative humidity levels on present-day mars.
  Icarus 333, 481 -- 495.
\newline\urlprefix\url{http://www.sciencedirect.com/science/article/pii/S0019103518305529}

\bibitem[{Renno et~al.(2009)Renno, Bos, Catling, Clark, Drube, Fisher, Goetz,
  Hviid, Keller, Kok, Kounaves, Leer, Lemmon, Madsen, Markiewicz, Marshall,
  McKay, Mehta, Smith, Zorzano, Smith, Stoker, and Young}]{renno2009}
Renno, N., Bos, B., Catling, D., Clark, B., Drube, L., Fisher, D., Goetz, W.,
  Hviid, S., Keller, H., Kok, J., Kounaves, S., Leer, K., Lemmon, M., Madsen,
  M., Markiewicz, W., Marshall, J., McKay, C., Mehta, M., Smith, M., Zorzano,
  M., Smith, P., Stoker, C., Young, S., 10 2009. Possible physical and
  thermodynamical evidence for liquid water at the phoenix landing site.
  Journal of Geophysical Research: Space Physics 114~(10).

\bibitem[{{Riu} et~al.(2018){Riu}, {Poulet}, {Gondet}, {Langevin}, {Carter},
  and {Bibring}}]{riu2018}
{Riu}, L., {Poulet}, F., {Gondet}, B., {Langevin}, Y., {Carter}, J., {Bibring},
  J.-P., Sep. 2018. {Global distribution of mafic minerals abundances and
  associated chemical composition at Mars: a legacy of OMEGA}. European
  Planetary Science Congress 12, EPSC2018--561.

\bibitem[{{Rivera-Valentin} and {Chevrier}(2015)}]{riveravalentin2015}
{Rivera-Valentin}, E.~G., {Chevrier}, V.~F., Jun 2015. {Revisiting the Phoenix
  TECP data: Implications for regolith control of near-surface humidity on
  Mars}. Icarus 253, 156--158.

\bibitem[{{Rivera-Valentín} et~al.(2018){Rivera-Valentín}, {Gough},
  {Chevrier}, {Primm}, {Martínez}, and {Tolbert}}]{rivera2018}
{Rivera-Valentín}, E.~G., {Gough}, R.~V., {Chevrier}, V.~F., {Primm}, K.~M.,
  {Martínez}, G.~M., {Tolbert}, M., 2018. Constraining the potential liquid
  water environment at gale crater, mars. Journal of Geophysical Research:
  Planets 123~(5), 1156--1167.
\newline\urlprefix\url{https://agupubs.onlinelibrary.wiley.com/doi/abs/10.1002/2018JE005558}

\bibitem[{{Savij{\"a}rvi} et~al.(2019){Savij{\"a}rvi}, {McConnochie}, {Harri},
  and {Paton}}]{savi2019}
{Savij{\"a}rvi}, H., {McConnochie}, T.~H., {Harri}, A.-M., {Paton}, M., Jul
  2019. {Water vapor mixing ratios and air temperatures for three martian years
  from Curiosity}. Icarus 326, 170--175.

\bibitem[{Savijärvi et~al.(2015)Savijärvi, Harri, and
  Kemppinen}]{savijarvi2015}
Savijärvi, H.~I., Harri, A.-M., Kemppinen, O., 2015. Mars science laboratory
  diurnal moisture observations and column simulations. Journal of Geophysical
  Research: Planets 120~(5), 1011--1021.
\newline\urlprefix\url{https://agupubs.onlinelibrary.wiley.com/doi/abs/10.1002/2014JE004732}

\bibitem[{{Smith}(2004)}]{smith2004}
{Smith}, M.~D., Jan. 2004. {Interannual variability in TES atmospheric
  observations of Mars during 1999-2003}. Icarus 167, 148--165.

\bibitem[{{Smith} et~al.(2009){Smith}, {Tamppari}, {Arvidson}, {Bass},
  {Blaney}, {Boynton}, {Carswell}, {Catling}, {Clark}, {Duck}, {DeJong},
  {Fisher}, {Goetz}, {Gunnlaugsson}, {Hecht}, {Hipkin}, {Hoffman}, {Hviid},
  {Keller}, {Kounaves}, {Lange}, {Lemmon}, {Madsen}, {Markiewicz}, {Marshall},
  {McKay}, {Mellon}, {Ming}, {Morris}, {Pike}, {Renno}, {Staufer}, {Stoker},
  {Taylor}, {Whiteway}, and {Zent}}]{smith2009}
{Smith}, P.~H., {Tamppari}, L.~K., {Arvidson}, R.~E., {Bass}, D., {Blaney}, D.,
  {Boynton}, W.~V., {Carswell}, A., {Catling}, D.~C., {Clark}, B.~C., {Duck},
  T., {DeJong}, E., {Fisher}, D., {Goetz}, W., {Gunnlaugsson}, H.~P., {Hecht},
  M.~H., {Hipkin}, V., {Hoffman}, J., {Hviid}, S.~F., {Keller}, H.~U.,
  {Kounaves}, S.~P., {Lange}, C.~F., {Lemmon}, M.~T., {Madsen}, M.~B.,
  {Markiewicz}, W.~J., {Marshall}, J., {McKay}, C.~P., {Mellon}, M.~T., {Ming},
  D.~W., {Morris}, R.~V., {Pike}, W.~T., {Renno}, N., {Staufer}, U., {Stoker},
  C., {Taylor}, P., {Whiteway}, J.~A., {Zent}, A.~P., Jul 2009. {H$_{2}$O at
  the Phoenix Landing Site}. Science 325, 58.

\bibitem[{{Spiga} et~al.(2018){Spiga}, {Banfield}, {Teanby}, {Forget}, {Lucas},
  {Kenda}, {Rodriguez Manfredi}, {Widmer-Schnidrig}, {Murdoch}, {Lemmon},
  {Garcia}, {Martire}, {Karatekin}, {Le Maistre}, {Van Hove}, {Dehant},
  {Lognonn{\'e}}, {Mueller}, {Lorenz}, {Mimoun}, {Rodriguez}, {Beucler},
  {Daubar}, {Golombek}, {Bertrand}, {Nishikawa}, {Millour}, {Rolland},
  {Brissaud}, {Kawamura}, {Mocquet}, {Martin}, {Clinton}, {Stutzmann}, {Spohn},
  {Smrekar}, and {Banerdt}}]{spiga2018}
{Spiga}, A., {Banfield}, D., {Teanby}, N.~A., {Forget}, F., {Lucas}, A.,
  {Kenda}, B., {Rodriguez Manfredi}, J.~A., {Widmer-Schnidrig}, R., {Murdoch},
  N., {Lemmon}, M.~T., {Garcia}, R.~F., {Martire}, L., {Karatekin}, {\"O}., {Le
  Maistre}, S., {Van Hove}, B., {Dehant}, V., {Lognonn{\'e}}, P., {Mueller},
  N., {Lorenz}, R., {Mimoun}, D., {Rodriguez}, S., {Beucler}, {\'E}., {Daubar},
  I., {Golombek}, M.~P., {Bertrand}, T., {Nishikawa}, Y., {Millour}, E.,
  {Rolland}, L., {Brissaud}, Q., {Kawamura}, T., {Mocquet}, A., {Martin}, R.,
  {Clinton}, J., {Stutzmann}, {\'E}., {Spohn}, T., {Smrekar}, S., {Banerdt},
  W.~B., Oct. 2018. {Atmospheric Science with InSight}. Space Science Reviews
  214, 109.

\bibitem[{{Szynkiewicz} et~al.(2009){Szynkiewicz}, {Ewing}, {Fishbaugh},
  {Bourke}, {Bustos}, and {Pratt}}]{szynkiewicz2009}
{Szynkiewicz}, A., {Ewing}, R.~C., {Fishbaugh}, K.~E., {Bourke}, M.~C.,
  {Bustos}, D., {Pratt}, L.~M., Mar. 2009. {Geomorphological Evidence of
  Plausible Water Activity and Evaporatic Deposition in Interdune Areas of the
  Gypsum-rich Olympia Undae Dune Field}. In: Lunar and Planetary Science
  Conference. Vol.~40 of Lunar and Planetary Science Conference. p. 2038.

\bibitem[{{Tanaka} et~al.(2014){Tanaka}, {Robbins}, {Fortezzo}, {Skinner}, and
  {Hare}}]{tanaka2014}
{Tanaka}, K.~L., {Robbins}, S.~J., {Fortezzo}, C.~M., {Skinner}, J.~A., {Hare},
  T.~M., May 2014. {The digital global geologic map of Mars:
  Chronostratigraphic ages, topographic and crater morphologic characteristics,
  and updated resurfacing history}. Planet. Space Sci. 95, 11--24.

\bibitem[{{Toner} et~al.(2014){Toner}, {Catling}, and {Light}}]{toner2014}
{Toner}, J.~D., {Catling}, D.~C., {Light}, B., May 2014. {The formation of
  supercooled brines, viscous liquids, and low-temperature perchlorate glasses
  in aqueous solutions relevant to Mars}. Icarus 233, 36--47.

\bibitem[{{Vaucher} et~al.(2009){Vaucher}, {Baratoux}, {Mangold}, {Pinet},
  {Kurita}, and {Gr{\'e}goire}}]{vaucher2009}
{Vaucher}, J., {Baratoux}, D., {Mangold}, N., {Pinet}, P., {Kurita}, K.,
  {Gr{\'e}goire}, M., Dec. 2009. {The volcanic history of central Elysium
  Planitia: Implications for martian magmatism}. Icarus 204, 418--442.

\bibitem[{{Velasco} and {Rodr{\'{\i}}guez-Manfredi}(2015)}]{velasco2015}
{Velasco}, T., {Rodr{\'{\i}}guez-Manfredi}, J.~A., Apr. 2015. {The TWINS
  Instrument On Board Mars Insight Mission}. In: EGU General Assembly
  Conference Abstracts. Vol.~17 of EGU General Assembly Conference Abstracts.
  p. 2571.

\bibitem[{{Warner} et~al.(2017){Warner}, {Golombek}, {Sweeney}, {Fergason},
  {Kirk}, and {Schwartz}}]{warner2017}
{Warner}, N.~H., {Golombek}, M.~P., {Sweeney}, J., {Fergason}, R., {Kirk}, R.,
  {Schwartz}, C., Oct. 2017. {Near Surface Stratigraphy and Regolith Production
  in Southwestern Elysium Planitia, Mars: Implications for Hesperian-Amazonian
  Terrains and the InSight Lander Mission}. Space Sci. Rev. 211, 147--190.

\bibitem[{{Yu} et~al.(2018){Yu}, {Russell}, {Rowe}, {Leneman}, {Lai}, {Cruce},
  {Means}, {Joy}, {Chi}, {Johnson}, {Mittelholz}, {Smrekar}, and
  {Banerdt}}]{yu2018}
{Yu}, Y., {Russell}, C.~T., {Rowe}, K., {Leneman}, D., {Lai}, H., {Cruce},
  P.~R., {Means}, J.~D., {Joy}, S.~P., {Chi}, P.~J., {Johnson}, C.,
  {Mittelholz}, A., {Smrekar}, S.~E., {Banerdt}, W.~B., Dec. 2018. {InSight
  Fluxgate Magnetometer: First Magnetic Measurements on the Mars Surface}. AGU
  Fall Meeting Abstracts.

\bibitem[{{Zent} et~al.(2009){Zent}, {Hecht}, {Cobos}, {Campbell}, {Campbell},
  {Cardell}, {Foote}, {Wood}, and {Mehta}}]{zent2009}
{Zent}, A.~P., {Hecht}, M.~H., {Cobos}, D.~R., {Campbell}, G.~S., {Campbell},
  C.~S., {Cardell}, G., {Foote}, M.~C., {Wood}, S.~E., {Mehta}, M., Mar 2009.
  {Thermal and Electrical Conductivity Probe (TECP) for Phoenix}. Journal of
  Geophysical Research (Planets) 114~(E3), E00A27.

\end{thebibliography}

\end{document}